\documentclass[9pt,twocolumn]{extarticle}
\usepackage[T1]{fontenc}
\usepackage[utf8]{inputenc}

\usepackage[paperwidth=595.276pt,paperheight=790.866pt,left=39pt,right=39pt,top=38pt,bottom=46pt,headheight=16pt,headsep=10pt,footskip=21pt]{geometry}
\usepackage{amsmath,mathtools,amsthm}
\usepackage{newtxtext,newtxmath}
\newtheorem{proposition}{Proposition}
\usepackage{graphicx,xcolor,microtype}
\definecolor{npjBlue}{HTML}{0065A3}
\usepackage{booktabs,tabularx,array,seqsplit,multirow}
\usepackage{caption}
\usepackage{float}
\usepackage{dblfloatfix}
\usepackage{fancyhdr}
\usepackage{titlesec}
\usepackage{url}
\usepackage[numbers,sort&compress,super]{natbib}
\usepackage{pdfpages}
\usepackage{hyperref}
\hypersetup{colorlinks=true,linkcolor=black,citecolor=npjBlue,urlcolor=npjBlue,
 pdftitle={AutoResearch: A Multi-Agent Framework for Reliable Research Workflow Automation},
 pdfauthor={Rajesh Kumar, Nikesh Lagun, Waqar Ali, Junaid Ahmed, Abdullah Aman Khan, Shaoning Zeng}}
\graphicspath{{figs/}}
\microtypesetup{protrusion=true,expansion=true}
\DeclareCaptionLabelSeparator{npjsep}{\space\textbar\space}

\titleformat{\section}{\sffamily\bfseries\fontsize{10.7}{12}\selectfont}{}{0pt}{}
\titlespacing*{\section}{0pt}{12pt plus 2pt minus 1pt}{5pt}
\titleformat{\subsection}{\sffamily\bfseries\fontsize{9.5}{11}\selectfont}{}{0pt}{}
\titlespacing*{\subsection}{0pt}{8pt plus 1pt minus 1pt}{3pt}
\titleformat{\subsubsection}{\sffamily\bfseries\fontsize{9}{11}\selectfont}{}{0pt}{}
\titlespacing*{\subsubsection}{0pt}{6pt plus 1pt minus 1pt}{2pt}

\newcommand{\suppl}[1]{Supplement~#1}
\fancypagestyle{firstpage}{%
 \fancyhf{}%
 \fancyfoot[R]{\sffamily\bfseries\fontsize{7.4}{8.4}\selectfont\thepage}%
}
\fancypagestyle{plain}{\pagestyle{fancy}}
\begin{document}
\thispagestyle{firstpage}
\twocolumn[{
\begin{minipage}{\textwidth}
\vspace{8pt}
\noindent\rule{\textwidth}{0.55pt}\par\vspace{8pt}
{\sffamily\bfseries\fontsize{21.5}{24.5}\selectfont AutoResearch: A Multi-Agent Framework for Reliable Research Workflow Automation\par}
\vspace{12pt}
\noindent\rule{\textwidth}{0.4pt}\par\vspace{3pt}
{\sffamily\bfseries\fontsize{8.7}{11}\selectfont
\href{https://orcid.org/0000-0003-0813-7485}{Rajesh Kumar}\textsuperscript{1}, \href{https://orcid.org/0009-0005-6372-4852}{Nikesh Lagun}\textsuperscript{2,3}, \href{https://orcid.org/0000-0003-0846-7281}{Waqar Ali}\textsuperscript{4}, Junaid Ahmed\textsuperscript{5}, Abdullah Aman Khan\textsuperscript{1} \& Shaoning Zeng\textsuperscript{6}\par}
\vspace{6pt}
\noindent\includegraphics[width=0.94\textwidth]{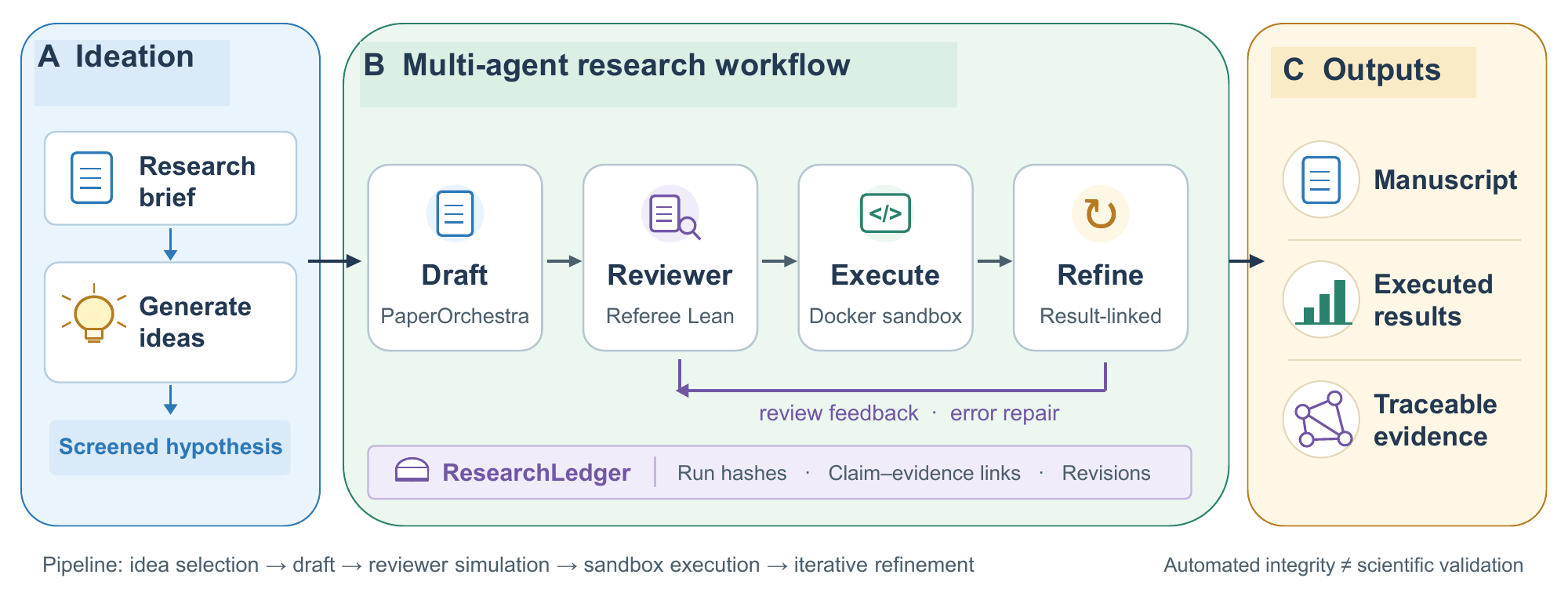}\par
\vspace{2pt}
\noindent{\sffamily\bfseries\fontsize{7.2}{9}\selectfont Graphical abstract}\quad
{\sffamily\fontsize{7.2}{9}\selectfont AutoResearch workflow from hypothesis generation to integrity-checked research artifacts.}\par
\vspace{8pt}
\vspace{6pt}\noindent\rule{\textwidth}{0.6pt}\par\vspace{7pt}
{\sffamily\fontsize{9.4}{12.2}\selectfont
Large language model (LLM)-based research agents can automate literature retrieval, code generation, experiment execution, and scientific writing. However, existing systems frequently produce non-executable code, unsupported claims, and unverifiable citations, limiting their reliability for practical research workflows. This study proposes AutoResearch, an execution-grounded multi-agent framework designed to improve the reliability, traceability, and automation of end-to-end research workflows through integrated execution validation, citation verification, and claim-support auditing. AutoResearch combines specialized agents for literature retrieval, code synthesis, iterative code repair, sandboxed Python/PyTorch execution, citation validation, claim-support assessment, workflow decision control, and structured \LaTeX{} artifact generation. The framework treats runtime execution failures, citation inconsistencies, and review-agent feedback as corrective signals to refine generated artifacts. The system was evaluated using HumanEval, MBPP, a SciCode subset, citation-validation tasks, claim-support auditing tasks, and end-to-end workflow stress-testing scenarios. Experimental evaluation demonstrates that AutoResearch achieves higher execution success rates, improved citation validity, stronger local claim support, and more reliable workflow completion compared with directly comparable baseline approaches. The framework consistently reduced unsupported outputs and improved the robustness of generated research artifacts across code-generation and scientific-assistance tasks. AutoResearch demonstrates that execution-grounded multi-agent orchestration can substantially improve the reliability of automated research assistance systems. The proposed framework provides a practical reliability-oriented research assistant that integrates executable validation and evidence-aware reasoning, while avoiding claims of fully autonomous scientific discovery. (Source Code: \url{https://github.com/raja21068/ResearchLedger})\par}
\vspace{15pt}
{\fontsize{7.2}{9.1}\selectfont\noindent
\textsuperscript{1} International Research Center for Complexity Sciences, Hangzhou International Innovation Institute, Beihang University, Hangzhou 311115, China.\quad
\textsuperscript{2} Hangzhou International Innovation Institute, Beihang University, Hangzhou, Zhejiang 311115, China.\quad
\textsuperscript{3} Lagun Lab, Kathmandu, Nepal.\quad
\textsuperscript{4} College of Science, Mathematics and Technology, Wenzhou-Kean University, Wenzhou, China.\quad
\textsuperscript{5} Computer Systems Engineering Department, Sukkur IBA University, Sindh, Pakistan.\quad
\textsuperscript{6} Yangtze Delta Region Institute (Huzhou), University of Electronic Science and Technology of China, Huzhou 313001, China.\par}
\vspace{10pt}
\noindent{\color{black!70}\rule{\textwidth}{0.4pt}}\par\vspace{6pt}
\end{minipage}
}]
\clearpage
\refstepcounter{section}
\label{sec:intro}
Research agents now search literature, formulate hypotheses, run code and draft manuscripts \cite{lu2026aiscientist,yamada2025aiscientistv2,schmidgall2025agentlab}. Paper2Agent extends this direction by converting papers, code, data and workflows into validated MCP-based executable agents that can reproduce analyses and answer new scientific queries \cite{miao2026paper2agent}. Recent work has also shifted attention from generation to \emph{verifiability}: ScientistOne maintains chains of evidence, EviGraph makes a typed evidence graph the agent's operational state, ARIS audits manuscript statements against a claim ledger, and claim-level provenance has been argued to be an evaluation target in its own right \cite{meng2026scientistone,ren2026evigraph,yang2026aris,rasheed2026auditability}. Agentification makes a remaining systems question more acute: after an upstream code, data, result or evidence correction, which persisted outputs, claims, decisions and manuscript statements remain current \emph{relative to their registered dependencies}? A corrected run can change one artifact while downstream state remains unchanged, even when both old and new software execute successfully. We call this a \emph{stale registered dependency}; the term concerns recorded state, not an independent judgement that the scientific conclusion is false.

\begin{figure*}[!t]
\centering
\includegraphics[width=\textwidth]{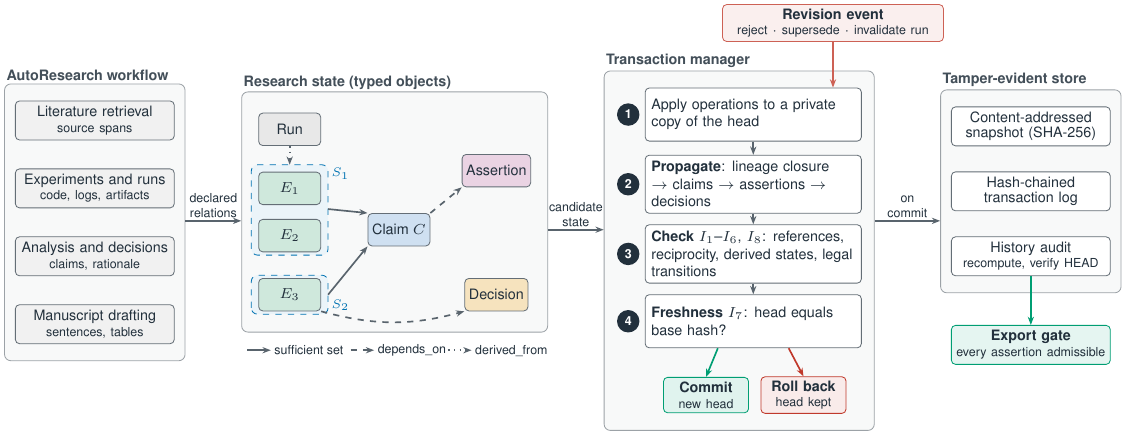}
\caption{TRL architecture. Pipeline stages register typed relations among runs, evidence, claims, decisions and manuscript assertions. A revision event enters the transaction manager, which applies it to a private copy, propagates it through the lineage closure, checks invariants $I_1$--$I_8$, and either commits a content-addressed, hash-chained snapshot or rolls back. The export gate admits a manuscript only if every assertion is ledger-admissible under the registered graph.}
\label{fig:framework}
\end{figure*}

We study this as a problem of \emph{revision semantics} over a typed, versioned \emph{registered research state} (Fig.~\ref{fig:framework}). The Transactional Research Ledger (TRL) applies each change to a private copy, propagates it through the registered dependency graph $G$, checks global invariants, and either commits a content-addressed snapshot or rolls back. Its contract is deliberately conditional: if a dependency is absent from or mistyped in $G$, the transaction layer cannot infer it. Two design decisions carry most of the weight. First, a claim's dependence on evidence is not generally a set of independent edges: a claim that needs an experiment \emph{and} a robustness check survives the loss of one only if a different sufficient set remains. TRL therefore represents claims by \emph{support sets}. Second, the label ``verified'' is separated into what deterministic code can check (artifact, execution, result) and what it cannot (scientific validity), so that TRL never certifies more than it can inspect.

We do not claim novelty for evidence graphs, provenance, truth maintenance, optimistic concurrency or transactional agent state, nor for support sets themselves, which correspond to the monotone fragment of ATMS labels and positive-Boolean provenance (Section~\ref{sec:related}). The contribution is a research-specific \emph{composition and operational contract}: typed recorded commitments, AND/OR dependency semantics, deterministic propagation, transactional activation, manuscript gating, and explicit measurement of dependency-registration quality. TRL is therefore an execution engine for supplied dependency declarations, not a method for discovering scientifically correct dependencies. The contributions are:
\begin{itemize}
    \item \textbf{Selective revision semantics}: support sets, required premises, non-epistemic dependencies and lineage represent conjunction, alternatives and transitive derivation beyond flat singleton support edges (Section~\ref{sec:support}).
    \item \textbf{Transactional integrity}: typed lifecycles and invariants $I_1$--$I_8$ provide atomic commit/rollback, predecessor freshness and tamper-evident local history, with guarantees explicitly scoped to the registered graph (Section~\ref{sec:txn}).
    \item \textbf{Operational construct boundaries}: provenance verification is separated from scientific validity, and ledger-liveness/admissibility are defined as graph-relative workflow states rather than truth values (Section~\ref{sec:states}).
    \item \textbf{Registration diagnostics}: typed precision, recall and support-set coverage distinguish missing, flattened and spurious relations, making the principal unprotected failure mode measurable (Sections~\ref{sec:registration}, \ref{sec:reg_results}).
    \item \textbf{Layered validation}: synthetic mechanism/conformance experiments, 13 genuine historical revisions passed through a developer-authored tracked proxy, and separate transaction fault injection test different parts of the system without treating any of them as independent end-to-end scientific validation (Sections~\ref{sec:design}--\ref{sec:hist_replay_main}).
    \item \textbf{A preregistered external-validation instrument}: the remaining confirmatory study uses native old/new executable-paper-agent trajectories, blind annotation, matched logical baselines and repository-clustered uncertainty (Section~\ref{sec:design}).
\end{itemize}
Our questions are: \textbf{RQ1} does TRL preserve invariants and prevent stale claims \emph{given a registered graph}; \textbf{RQ2} which mechanisms are necessary, including support sets and lineage; \textbf{RQ3} is its state machine-checkable (deterministic validation, rollback, tamper-evident history) and practical up to $10^5$ triples; \textbf{RQ4} how sensitive is protection to registration quality; \textbf{RQ5} can genuine historical semantic revisions be carried through immutable runs, evidence supersession, claim invalidation, manuscript audit and recovery in the implemented ledger; and \textbf{RQ6} how well do independently generated real paper agents register dependencies and preserve scientific state under genuine revisions. RQ1--RQ4 are answered with mechanism experiments, RQ5 with a developer-run tracked-proxy bridge study, and RQ6 and all human-audit claims remain open. Extended tables, protocols and proofs are in the Supplement (\suppl{S1}--S10).

\subsection{Prior work and novelty boundary}
\label{sec:related}
\paragraph{Research agents and evidence assurance}
The AI Scientist, its tree-search successor and Agent Laboratory connect ideation, experimentation and writing \cite{lu2026aiscientist,yamada2025aiscientistv2,schmidgall2025agentlab}. Paper2Agent converts a paper and its public computational assets into tested MCP tools and reports robustness to several execution-level repository defects \cite{miao2026paper2agent}. That addresses whether a paper agent can be made executable and kept available; our target is complementary: whether \emph{persisted downstream research state} remains admissible after a scientifically consequential upstream revision, including a revision for which both versions still execute. ScientistOne requires generated claims to stay traceable to evidence across the lifecycle; EviGraph repairs downstream subgraphs when a weak dependency is found; ARIS audits manuscript statements against a claim ledger and raw evidence \cite{meng2026scientistone,ren2026evigraph,yang2026aris}. Auditability itself has been proposed as an evaluation dimension \cite{rasheed2026auditability,martinboyle2026papertrail}; we do not evaluate it, as we ran no human study. None of these was run head-to-head with TRL; \suppl{S8} tabulates their reported mechanisms.

\paragraph{Truth maintenance, belief revision and argumentation}
Justification-based truth maintenance labels nodes IN or OUT from justifications and revises dependents when an assumption is withdrawn \cite{doyle1979tms}; assumption-based maintenance instead labels each node with the minimal sets of assumptions (environments) under which it holds and answers queries in any context \cite{dekleer1986atms}. TRL's claim liveness is the monotone fragment of both: a sufficient set is an in-list, required premises and non-epistemic dependencies are in-list members common to every justification, lineage is the justification of derived evidence, and a claim's family of sufficient sets is its ATMS label. The same function is the positive-Boolean provenance polynomial of the claim over its evidence, and rejecting evidence is deletion propagation \cite{buneman2001provenance,green2007provenance}. We test this correspondence (Section~\ref{sec:tms_xc}, \suppl{S8}). TRL lacks non-monotonic (OUT) justifications, nogood-directed backtracking and an ATMS's lattice of contexts; contradicting evidence only makes a claim \texttt{contested}. What it adds is orthogonal: atomic, predecessor-anchored commit and rollback, hash-chained history, typed lifecycles and a fail-closed export gate. AGM theory characterizes rational revision \cite{alchourron1985agm}; structured argumentation represents claims as conclusions of premise sets under attack relations \cite{toulmin1958uses,dung1995acceptability,modgil2014aspic}. Our support sets are a restricted, \emph{declared} fragment: sufficient sets are premise sets, required premises are common to all of them, contradicting evidence is a rebutter. TRL neither weighs argument strength nor revises beliefs; it revises recorded commitments.

\paragraph{Provenance, workflows and research objects}
Database provenance and W3C PROV model derivation \cite{buneman2001provenance,green2007provenance,w3c2013prov}; scientific-workflow systems distinguish prospective from retrospective provenance \cite{davidson2008provenance,simmhan2005survey}; Research Objects, RO-Crate, ReproZip and FAIR package artifacts for reuse \cite{bechhofer2013linked,soilandreyes2022rocrate,chirigati2016reprozip,wilkinson2016fair}. They describe a package; TRL maintains its \emph{state} under revision. A derivation record cannot say whether an input was necessary, sufficient or incidental to a claim; support sets can, but are declared, not captured.

\paragraph{Transactions and propagation}
TRL's propagation is a restricted instance of incremental view maintenance \cite{gupta1995maintenance}, its freshness invariant an application of optimistic concurrency control \cite{kung1981optimistic}, and predecessor-anchored activation is shared with the Continuity Kernel for generic agent state \cite{he2026continuity}. Build systems and self-adjusting computation likewise invalidate and recompute dependents of a changed input \cite{feldman1979make,mokhov2018build,acar2009selfadjusting}. The difference is the object being maintained: TRL revises recorded commitments (claims, decisions and manuscript assertions) whose dependency semantics are declared and may include alternative sufficient paths, rather than merely recomputing a derived value. This is an architectural distinction, not a claim that invalidation, transactions or versioned state are new.

\paragraph{Novelty boundary}
The closest formal account of TRL's support logic is ATMS/positive-Boolean provenance; the closest systems ingredients are provenance/workflow lineage, dependency invalidation and predecessor-anchored transactional activation. Accordingly, our novelty claim is limited to the \emph{combination} of these mechanisms for typed research commitments and to its registration-aware evaluation. Table~S11 compares the reported mechanisms of the closest research-agent systems, and Table~S12 gives the exact correspondence to truth-maintenance/provenance models. ``Not reported'' in that comparison is descriptive, not evidence that another implementation could not support the feature. The survey is not exhaustive of all event-sourcing, data-versioning or ML-pipeline-lineage systems, so we make no universal absence claim outside the systems and formalisms analyzed here.

\section{Results}
\label{sec:results}
\subsection{Conformance benchmarks reproduce}
Table~\ref{tab:conf} summarizes the earlier benchmarks. Their result files are identical to the previous prototype's apart from timing fields; claim states now use the typed vocabulary, which changes no count. These results establish that the implementation matches its specification, not that the specification captures real revision. DRS-1400 makes one boundary explicit: when a true dependency is never registered, TRL raises no violation and ledger-reported escape stays at 0\%, while ground-truth escape equals the unregistration rate.

\begin{table*}[!t]
\centering
\caption{Conformance benchmarks on the revised core (full tables in \suppl{S3}). Residual violations and escape are for the state after all mutations.}
\label{tab:conf}
\footnotesize
\begin{tabularx}{\columnwidth}{l>{\raggedright\arraybackslash}X>{\raggedright\arraybackslash}X}
\toprule
Benchmark & Full TRL & Strongest non-TRL condition \\
\midrule
RSM-1000 & 0 violations, escape 0\%, all legitimate updates committed, all corrupt ones rejected & post-hoc repair: 1.206 violations/trial; raw state: escape 79.9\% \\
RSM-SEQ-2500 & 100/100 episodes consistent & post-hoc repair: 30.0 violations at episode end \\
RSR-500 & OCC + rebase: both updates kept in 500/500 & last-write-wins: lost update in 500/500 \\
HT-250 & 250/250 corruptions detected, 0/25 false positives & --- \\
RLV2-400 & 350/350 corruptions rejected with target rule, 50/50 clean accepted & --- \\
DRS-1400 & ledger escape 0\% at every rate; ground-truth escape $=p_{\text{missing}}$ & --- \\
\bottomrule
\end{tabularx}
\end{table*}

\subsection{Support sets and lineage (SUP-300)}
\label{sec:sup_results}
\begin{figure*}[!t]
\centering
\includegraphics[width=\textwidth]{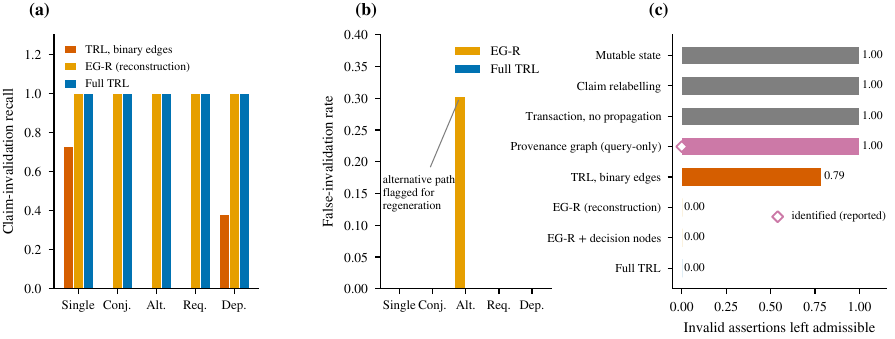}
\caption{SUP-300 on synthetic trajectories with perfect registration (815 events). (a) Claim-invalidation recall by claim template: single, conjunctive, alternative-group, required-premise, non-epistemic dependency. (b) False invalidation of still-justified claims: only the reconstruction of EviGraph's mechanism over-invalidates, and only where an alternative sufficient path exists. (c) Invalid assertions left admissible, by condition; the diamond marks what the query-only provenance graph \emph{reports} rather than updates.}
\label{fig:sup}
\end{figure*}

\emph{Synthetic fixtures with generator-authored gold; magnitudes depend on the arbitrary template mix.} With perfect registration, full TRL's set of invalidated claims equalled the oracle's in all 815 events, and it recovered every affected decision and assertion (Fig.~\ref{fig:sup}, Table~\ref{tab:sup}). This is specification-level agreement on generator-authored graphs. It supports implementation correctness and the expressiveness claim for AND/OR dependencies, but it is not an estimate of real-world impact-identification accuracy.

These are representation sanity checks, not comparative evidence: the generator includes templates that singleton edges cannot express, so their failure is by construction. The other conditions fail in ways that map to semantic features. Claim relabelling and TRL without support sets recover 20.5\% of invalidated claims: none of the conjunctive, alternative-group or required-premise claims, and 72.8\% of singleton claims (the rest are lineage taint). A provenance store without transactions \emph{identifies} every affected object but updates none, so all invalid assertions remain admissible (escape 1.00 by state, 0.00 by report). The EviGraph reconstruction recovers every affected claim and assertion but flags 30.3\% of still-justified claims that have an alternative sufficient path, because flagging a whole subgraph cannot tell a lost path from a lost claim; its decision handling depends on whether decisions are graph nodes. Its over-invalidation costs needless regeneration rather than unsound state, and is favored by the given-invalid-node assumption. EG-R is a synthetic reachability baseline: the original system is reported to include further components (for example semantic consistency checks and graph checkpointing) that are not modelled, and no TRL-versus-EviGraph conclusion follows.

\begin{table*}[!t]
\centering
\caption{SUP-300 by condition. Recall: invalidated claims recovered; FIR: still-justified claims wrongly invalidated; Escape: invalid assertions left admissible. Values are over stored state unless marked ``reported''.}
\label{tab:sup}
\scriptsize
\begin{tabular}{@{}lcccc@{}}
\toprule
Condition & Recall & FIR & Dec.\ recall & Escape \\
\midrule
Mutable state & 0.000 & 0.000 & 0.000 & 1.000 \\
Claim relabelling & 0.205 & 0.000 & 0.000 & 1.000 \\
Provenance graph, updated & 0.000 & 0.000 & 0.000 & 1.000 \\
Provenance graph, reported & 1.000 & 0.070 & 1.000 & 0.000 \\
Transaction, no propagation & 0.000 & 0.000 & 0.000 & 1.000 \\
EG-R (reconstruction) & 1.000 & 0.070 & 0.000 & 0.000 \\
TRL, binary edges & 0.205 & 0.000 & 0.757 & 0.787 \\
\textbf{Full TRL} & \textbf{1.000} & \textbf{0.000} & \textbf{1.000} & \textbf{0.000} \\
\bottomrule
\end{tabular}
\end{table*}

\subsection{Registration sensitivity (REG)}
\label{sec:reg_results}
\begin{figure*}[!t]
\centering
\includegraphics[width=\textwidth]{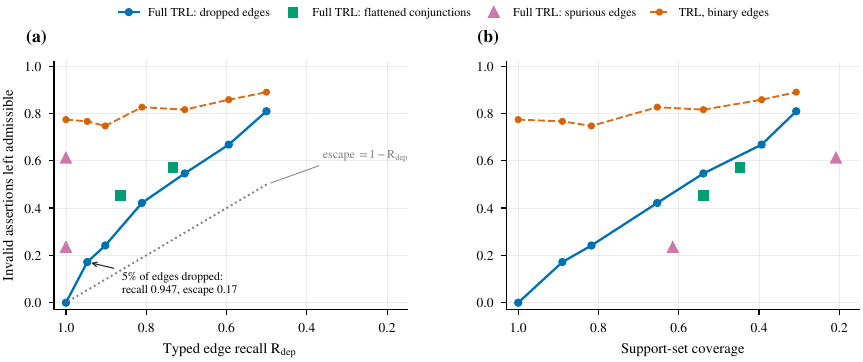}
\caption{Protection under registration corruption (100 synthetic trajectories per point). Escape is the fraction of invalid manuscript assertions left admissible. (a) Against typed edge recall; the dotted line is escape $=1-\mathrm{R}_{\mathrm{dep}}$. (b) Against support-set coverage. Dropped edges, flattened conjunctions and spurious edges are separate corruption processes; the dashed line is TRL restricted to binary edges.}
\label{fig:reg}
\end{figure*}

Fig.~\ref{fig:reg} shows escape of full TRL as the registered graph is corrupted; all observations are on synthetic fixtures. \emph{Escape rises much faster than edge recall falls}: dropping 5\% of true edges gives $\mathrm{R}_{\mathrm{dep}}=0.947$ and support-set coverage 0.890, yet 17.2\% of invalidated assertions escape; at 10\% dropped, recall is 0.902 and escape 24.2\%. Losing one member of a conjunction removes the whole claim from protection, and one lost lineage edge can do the same downstream. Over the seven drop rates escape $=0.074+1.52\,(1-\mathrm{R}_{\mathrm{dep}})$ ($R^2=0.977$, aggregate points, $n=7$; a description of these fixtures, not a law), which is why edge recall alone is an insufficient registration metric. \emph{Flattening} produces no missing edge yet yields 57.1\% escape when every conjunction is flattened. \emph{Spurious} edges with perfect recall still raise escape (23.4\% at 10\%, 61.2\% at 30\%): an extra \texttt{supports} edge creates a fake alternative path that keeps an invalid claim alive, so registration precision matters for protection, not only recall. TRL without support sets fails at every point.

\subsection{State-machine conformance (SM-CONF)}
The exhaustive matrix agreed with the documented relations on all 107 ordered pairs. In 150 fuzzing episodes (seed 53), 4,582 of 4,611 operations committed with no invariant violation, \emph{none} of the legitimate operations was rejected as an illegal transition, and the 29 rejections were all the intended \texttt{stale\_decision}. The fuzzer exercised 44 of 48 table edges; all 500 forbidden direct writes and 100 direct writes contradicting the derivation were rejected. The informative outcome is what the first pass did to the design: our initial claim machine declared \texttt{stale} sticky and forbade \texttt{supported}$\to$\texttt{unsupported}, but re-supporting a stale claim with unaccepted evidence and declaring an unaccepted required premise reach exactly those transitions. We corrected the specification: claim states are derived and the operative constraint is agreement with the derivation. SM-CONF tests consistency of the implementation with its own tables, not that they are the right tables for any workflow.

\subsection{Truth-maintenance cross-check (TMS-XC)}
\label{sec:tms_xc}
On 300 synthetic trajectories (seed 61; 900 events per graph) we recomputed each revision's affected objects with the oracle, a Doyle-style JTMS, a de Kleer-style ATMS and full TRL, on the gold graph and on graphs with dropped, flattened or spurious edges. JTMS and ATMS matched the oracle on all 3,600 events; TRL matched on 3,551. Its 49 divergences all have mechanical causes (\suppl{S8}): 25 are assertions already blocked by an unsupported claim, which undergo no state transition, and 24 are mistyped duplicate edges on one (evidence, claim) pair, which collide in TRL's one-relation-per-pair model and make it over-invalidate. This is agreement among implementations of one semantics, not evidence about real research; a relation model with several relations per pair would remove the collision.

\subsection{Scaling}
\label{sec:scale_results}
\begin{figure*}[!t]
\centering
\includegraphics[width=\textwidth]{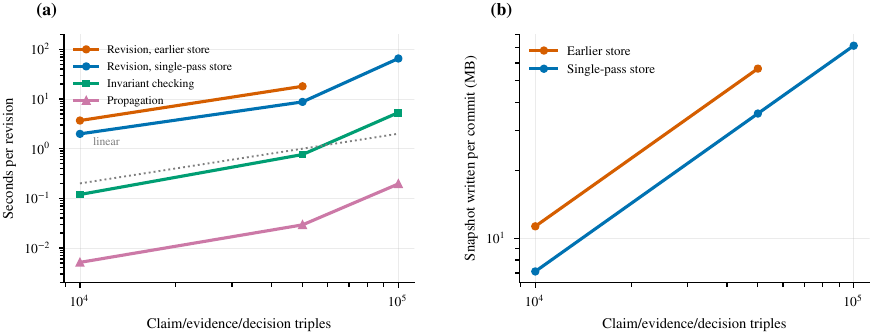}
\caption{Scaling of a single evidence-rejection revision, $10^4$--$10^5$ triples (median of 3--5). (a) Time per revision, invariant checking and propagation under the earlier and the single-pass store; the dotted line has unit log-log slope. (b) Snapshot size, which is also the number of bytes written per commit.}
\label{fig:scale}
\end{figure*}

Fig.~\ref{fig:scale} shows one revision on a committed archive of $10^4$ to $10^5$ triples (three records each). Timings are from a noisy 24-thread Windows laptop (the earlier prototype's code is $\approx\!3.5\times$ slower here than its published numbers, and medians at $10^3$ triples varied by up to $3\times$), so only scales $\ge10^4$ are interpreted. Three findings are robust. \emph{Propagation is not the bottleneck}: 5--30 ms through $5\times10^4$ triples and 0.2 s at $10^5$, under 1\% of a revision. \emph{Everything else is linear in archive size, and the full-snapshot design is the bottleneck}: a revision takes about 2 s at $10^4$ triples, 9--18 s at $5\times10^4$ and 65 s at $10^5$, with a 72 MB snapshot rewritten per commit. \emph{Storage flushing is not the cost at scale}: with \texttt{fsync} disabled, commit time at $\ge5\times10^4$ triples is unchanged. A single-pass store (one serialization, no deep copy) keeps hashes identical, roughly halves revision time and cuts snapshots by 37\%, but changes constants, not asymptotics. Density matters less than record count (at $10^4$ triples, 1 to 20 links per claim raises a revision from 2.0 to 2.5 s). We did not run $5\times10^5$ triples; supporting that regime needs a persistent Merkle or delta structure with lazy loading and incremental validation, which we describe but do not implement. Full tables are in \suppl{S6}.

\subsection{Historical revisions, tracked proxy trajectories and fault injection (development evidence)}
\label{sec:hist_replay_main}
The source audit identified 13 genuine semantic revisions across nine scientific-software repositories (\texttt{POP-TOOLS}, \texttt{scanpy}, \texttt{squidpy}, \texttt{TabPFN}, \texttt{SAELens}, \texttt{Compass}, \texttt{scvi-tools}, \texttt{seurat}, \texttt{scikit-learn}). Replaying the smallest changed semantic unit reproduced the documented old/new behavioral, numerical, data or configuration difference in 13/13 cases (\suppl{S9}). The cases include sparse normalization, a statistical-variance correction, calibration under majority downsampling, feature subsampling, sequence-boundary handling, PCA setup, duplicated data removal and scientific defaults. These are narrow semantic replays rather than full historical repository builds.

We then passed every case through a deterministic paper-agent proxy as two separate immutable ResearchLedger runs. The old output was frozen as verified bridge evidence for a claim; the historical revision superseded that evidence transactionally; and the revised output was attached as replacement evidence and verified. All 13 trajectories produced different old/new artifact hashes and completed the same expected ResearchLedger-native lifecycle (Table~\ref{tab:tracked_proxy}): the claim was \texttt{supported} before revision, became \texttt{hypothesis} after the old evidence was superseded, and returned to \texttt{supported} after replacement verification. As noted in Section~\ref{sec:design}, these are bridge-layer labels: \texttt{hypothesis} denotes absence of a currently sufficient live support path and should not be confused with the core TRL labels in Table~\ref{tab:states}. An explicit manuscript provenance marker naming the superseded evidence was flagged stale in 13/13 cases and recovered in 13/13 after the marker was changed to the replacement evidence. Every final workspace had zero validation errors. Median local transaction-only time to the safe invalidated state was 9.6 ms and median recovery bookkeeping after the revised output already existed was 15.0 ms; old/new proxy tool runs were about 2.9--3.0 s and were dominated by local process/environment capture. These timings are implementation diagnostics, not Paper2Agent or production benchmarks.

\begin{table*}[!t]
\centering
\caption{Ledger-integrated tracked proxy trajectories over 13 genuine historical revisions. Bindings and proxy tools were developer-authored; this is bridge/conformance evidence, not independent external validation.}
\label{tab:tracked_proxy}
\footnotesize
\begin{tabularx}{\columnwidth}{>{\raggedright\arraybackslash}Xr}
\toprule
Outcome & Result \\
\midrule
Historical revisions / repositories & 13 / 9 \\
Documented old/new semantic change reproduced & 13/13 \\
Old/new artifact hash changed & 13/13 \\
Claim \texttt{supported}$\to$\texttt{hypothesis} after supersession & 13/13 \\
Explicit superseded-evidence manuscript marker flagged stale & 13/13 \\
Replacement evidence restored \texttt{supported} claim & 13/13 \\
Recovered manuscript marker no longer stale & 13/13 \\
Final workspace validation errors & 0 in 13/13 \\
Median time to safe invalidated state$^\dagger$ & 9.6 ms \\
Median recovery bookkeeping$^\dagger$ & 15.0 ms \\
\bottomrule
\multicolumn{2}{l}{\scriptsize $^\dagger$Excludes generation of the revised tool output; local proxy diagnostic only.}
\end{tabularx}
\end{table*}

We separately fault-injected the multi-file revision transaction under six deterministic checks. All passed, including forced write failure and simulated mid-transaction crash with exact pre-image restoration, 11/11 tested entity-write fault positions on an 11-file revision, preservation of a claim when one of two alternative sufficient support paths remained live, and two independent concurrent revision processes both reaching the expected final state. On a separate authored 10-event structural pilot, support-set provenance without transactions and full TRL both reached 10/10 exact impact identification; flat binary edges reached 6/10; unconditional downstream reachability and invalidate-all/full rebuild reached 4/10 exact matches because they over-invalidated claims retaining an alternative sufficient support path. A repository-clustered bootstrap over eight repository clusters gave a median TRL-minus-binary-edge claim-recall difference of 0.40 (95\% percentile interval $[0.111,0.700]$) and a median TRL-minus-support-set-provenance exact-match difference of 0.0 ($[0,0]$). Because the impact graphs and gold sets were authored for mechanism development, these intervals are descriptive sensitivity analyses, not external-validation uncertainty. The decomposition is important: AND/OR support semantics explain selective impact identification on this pilot; the transaction layer contributes atomic multi-file state change, rollback, crash recovery, freshness/serialization and auditable history. The integrated validation package passed 206/206 automated tests across 22 modules.

\section{Discussion}
\label{sec:discussion}
\subsection{Implications}
The current evidence supports a narrower decomposition than a generic claim of ``research reliability.'' \emph{Artifact reliability} concerns whether code executes, artifacts match and recorded results can be reproduced; \emph{registered-state revision integrity} concerns whether a declared dependency change is reflected consistently in persisted claims, decisions and manuscript bindings. Executable-paper systems such as Paper2Agent make the distinction operational: a tool may remain executable while a semantic correction changes the output on which persisted claims depend. The tracked-proxy study shows that the implemented bridge can carry such an old/new change through immutable runs, supersession, explicit stale-evidence detection and recovery, but it does not establish that a native paper agent will register the correct dependency structure. The experiments further separate two mechanisms. AND/OR support semantics determine \emph{which} objects should change under the declared graph; the transaction layer determines whether that multi-object change is applied atomically and audibly. Registration sensitivity shows that neither mechanism protects an omitted or mistyped dependency. Consequently, precision/recall should be reported together with support-set coverage, and any claim about manuscript protection must also report assertion-binding quality. Whether deterministic gating improves human review or native agent behavior remains an empirical question, not a consequence of the state model.

\subsection{Threats to validity and limitations}
\begin{itemize}
    \item \textbf{Construct and external validity: no independent real-paper-agent result.} Synthetic experiments use generator-authored reference structures. The 13 historical revision identities and changed semantics are genuine, and the tracked proxy executes the ledger invalidation/recovery lifecycle, but the executable is a deterministic paper-agent proxy rather than a Paper2Agent-generated MCP and the claim/evidence bindings are developer-authored rather than blind human labels. Thus the study does not establish how often real agents register the right dependencies, how native agent outputs change across revisions, or how well the proposed relation vocabulary agrees with independent scientific judgement.
    \item \textbf{Construct assignment.} The machine semantics are operational once $\mathcal{S}(c)$, $R(c)$, $D(c)$, lineage and assertion bindings are supplied; deciding those relations is not machine-validated here. In particular, the distinction between a required premise and a non-epistemic dependency is an annotation decision whose inter-rater reliability remains unmeasured.
    \item \textbf{Semantic warrant and scope.} A support set is a Boolean declaration checked for structure, and provenance-verified does not mean well designed or validly inferred. The schema has no graded support/contradiction, evidence-quality model or representation of statistical dependence between apparently independent support paths. Claim-to-claim and decision-to-claim dependency are absent, and semantic claim edits currently require a new claim record with relations re-established (\suppl{S7}).
    \item \textbf{Comparator validity.} EG-R is a reconstruction of published mechanism; no original EviGraph system was run head-to-head, and no conclusion about the native system follows. JTMS/ATMS comparisons test correspondence to known semantics rather than competitive novelty.
    \item \textbf{Auditability.} No person has used TRL in the reported experiments. We claim deterministic checking of invariants, digests, bindings and the export gate and tamper evidence against local edits; we do not claim that reviewers find stale evidence faster or miss fewer stale statements (\suppl{S7} is a protocol only).
    \item \textbf{Registration and binding acquisition.} Real registration quality and assertion-binding quality are not measured, and automatic relation proposal from manuscript text or run logs is not implemented. These are upstream acquisition problems that bound achievable protection.
    \item \textbf{Internal and statistical validity.} Diagnostics use fixed seeds and one generator configuration. The 10-event historical structural pilot is authored development evidence; its repository-clustered interval quantifies sensitivity within that pilot, not population-level uncertainty. Timing is from one local machine, and the millisecond tracked-proxy transaction times exclude external agent generation and revised computation.
    \item \textbf{Scale and durability.} Scale is measured to $10^5$ triples using full snapshots. Larger or distributed deployments would require a persistent delta/Merkle design and stronger durability/failure models; the reported transaction tests are local, not distributed consensus or storage-device durability tests.
    \item \textbf{Novelty boundary.} Truth maintenance, provenance, argumentation, dependency invalidation and predecessor-anchored activation already exist \cite{doyle1979tms,buneman2001provenance,dung1995acceptability,feldman1979make,he2026continuity}. Support sets are a restricted declared fragment, not an advance on ATMS or Boolean provenance. The claimed contribution is the research-specific composition, operational boundary and registration-aware evaluation, and no universal claim is made over unsurveyed event-sourcing, data-versioning or ML-lineage systems.
\end{itemize}
Because a provenance layer lowers the cost of tracing but not the cost of judging, it should support, not replace, human review of generated manuscripts (\suppl{S9}).

\subsection{Conclusion}
We presented a transactional state-control layer for declared research dependencies. Its formal and implementation contribution is conditional but concrete: given a registered graph, AND/OR support semantics determine graph-relative liveness, typed invariants constrain legal state, and a predecessor-anchored transaction applies the resulting multi-object revision atomically or not at all. Synthetic mechanism experiments support conformance to that specification and show, within fixtures containing conjunctions and alternatives, why flat singleton edges are insufficient; registration diagnostics show that missing, flattened and spurious relations can each defeat protection. Thirteen genuine historical software revisions were additionally carried through a developer-authored immutable-run proxy lifecycle, and local fault injection exercised rollback and crash recovery. These results establish a mature implementation and a bridge from real semantic changes to persisted revision state, but they do not validate the upstream scientific dependency graph. The remaining publication-critical question is therefore construct and external validity: whether independent experts can reproducibly identify the proposed relations and whether native executable paper agents preserve state more safely under genuine revision. The preregistered old/new study is designed to answer that question with at least 60 revision events from at least 15 repositories, blind annotation, adjudication, strong logical baselines and repository-clustered uncertainty.

\section{Methods}
\subsection{Transactional research state}
\label{sec:trl_method}

\subsubsection{Research state and typed objects}
\label{sec:states}
\emph{Registered research state} is the set of recorded commitments---runs $U$, evidence records $E$, claims $C$, decisions $D$, manuscript assertions $A$---and the typed relations among them. It is not truth, belief probability or semantic warrant: TRL maintains the consistency of a \emph{declared} structure under revision and does not judge whether the declaration is complete or scientifically justified. Throughout the paper, \emph{live} and \emph{admissible} mean \emph{ledger-live} and \emph{ledger-admissible with respect to the registered graph $G$}. Each object kind has a closed state set and legal transitions (Table~\ref{tab:states}); a committed transition outside the relation is rejected (invariant $I_8$). Labels are workflow states, not truth values, and \texttt{rejected} (evidence) and \texttt{stale} (claim) are deliberately not the same thing.

\begin{table*}[!t]
\centering
\caption{Per-object state machines (full relations in \suppl{S1}). Claim and assertion states are derived from the support structure; the rest change by explicit operation.}
\label{tab:states}
\footnotesize
\begin{tabularx}{\columnwidth}{l>{\raggedright\arraybackslash}X>{\raggedright\arraybackslash}X}
\toprule
Object & States & Notes \\
\midrule
Evidence & proposed, accepted, provenance\_verified, rejected, superseded & \texttt{rejected} and \texttt{superseded} terminal: reviving a result needs a new record \\
Claim & unsupported, supported, verified, contested, stale & derived; stored state must equal the derivation ($I_2$) \\
Decision & proposed, decided, revisiting, revised, abandoned & decided/revised may not cite non-live evidence ($I_4$) \\
Assertion & draft, admissible, blocked, outdated & admissible/blocked derived from bound claims \\
Run & created, successful, failed, invalidated & failed, invalidated terminal \\
\bottomrule
\end{tabularx}
\end{table*}

\begin{table*}[!t]
\centering
\caption{Operational scope of the central constructs. The middle column is machine-checkable once records and relations are supplied; the right column is an external scientific judgement that TRL does not infer.}
\label{tab:constructs}
\footnotesize
\begin{tabularx}{\textwidth}{p{2.5cm}>{\raggedright\arraybackslash}X>{\raggedright\arraybackslash}X}
\toprule
Construct & Operational meaning inside TRL & Outside TRL / requires external judgement \\
\midrule
Evidence record & Versioned object with source/run/artifact identity and a workflow state & Whether the object is scientifically relevant or valid evidence for a proposition \\
Sufficient set $S\in\mathcal{S}(c)$ & All members must be ledger-live; one complete set is enough to keep $c$ live & Whether those records are in fact jointly sufficient for the scientific claim \\
Required premise $R(c)$ & Every member must remain ledger-live in addition to any sufficient set & Whether the item is genuinely necessary across all evidentiary routes \\
Non-epistemic dependency $D(c)$ & Loss forces re-examination/staleness but does not raise the provenance support level $\lambda$ & Whether a code, preprocessing, model or assumption change should invalidate the claim \\
Lineage & Declared \texttt{derived\_from} relation whose liveness propagates transitively & Whether unrecorded or latent dependence exists \\
Assertion binding & Explicit span-to-claim relation used by the export gate & Text segmentation, binding granularity and whether the assertion is scientifically warranted \\
Ledger-live / ledger-admissible & Deterministic status derived from $G$, object states and invariants & Truth, credibility, effect size, evidentiary strength or reviewer acceptance \\
\bottomrule
\end{tabularx}
\end{table*}

\paragraph{Provenance-verified evidence}
We separate four things: \emph{artifact verified} (declared artifacts exist and match their SHA-256), \emph{execution verified} (the bound run succeeded and recorded an environment digest), \emph{result verified} (a declared metric is reproduced from the recorded artifact by a supplied checker), and \emph{scientifically validated} (the experiment is valid evidence for the proposition). TRL evaluates the first three deterministically; a record is \texttt{provenance\_verified} only if no tier failed and at least one passed, and a failed re-check demotes it. The fourth is deliberately outside the data model: a matching hash and a clean exit code do not make an experiment a valid test of a hypothesis. Accordingly a \texttt{verified} claim means only that every member of its best sufficient set is provenance-verified; it should be read as \emph{provenance-complete}, not as true or well tested. Table~\ref{tab:constructs} makes the same boundary explicit for the remaining constructs. The implementation retains the concise labels \texttt{live} and \texttt{admissible}; in interpretation they are always graph-relative workflow labels, never judgements of credibility or scientific warrant.

\subsubsection{Support-set semantics}
\label{sec:support}
\begin{figure*}[!t]
\centering
\includegraphics[width=\textwidth]{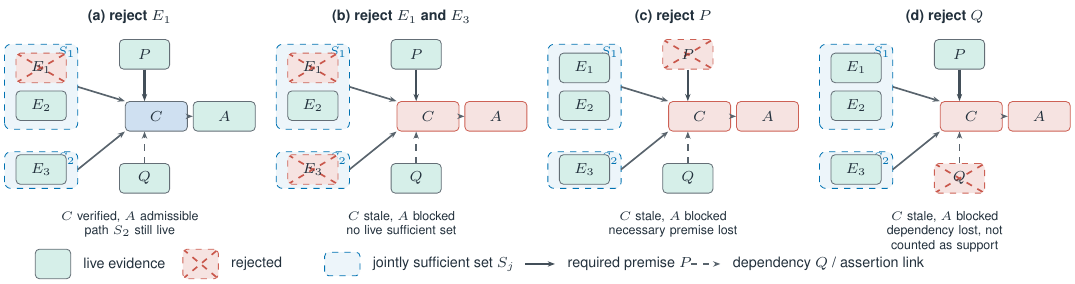}
\caption{Support-set semantics: one claim $C$ with sufficient sets $S_1=\{E_1,E_2\}$ and $S_2=\{E_3\}$, a necessary premise $P$, a non-epistemic dependency $Q$ and a manuscript assertion $A$, under four revisions. (a) Losing $E_1$ breaks $S_1$ but $S_2$ survives. (b) Losing $E_1$ and $E_3$ leaves no sufficient set. (c) Losing $P$ stales $C$ although both sets are intact. (d) Losing $Q$ stales $C$ without $Q$ ever counting as support. Under binary edges only outcome (a) is reproduced; in (b)--(d) $C$ and $A$ would be left unchanged.}
\label{fig:support}
\end{figure*}

Registering $E_1\!\to\! C$ and $E_2\!\to\! C$ as independent edges lets $C$ survive the loss of $E_1$ on the strength of $E_2$, even if $C$ needs both. We therefore derive for each claim $c$ a family $\mathcal{S}(c)$ of \emph{jointly sufficient} evidence sets, necessary premises $R(c)$, and non-epistemic dependencies $D(c)$ (evidence the claim depends on without it counting as support). Conjunction is a set with several members; disjunction is several sets. Let $\mathrm{live}(e)$ hold iff $e$ is neither rejected nor superseded, was not produced by a failed or invalidated run, and every record in its \texttt{derived\_from} lineage is live. Then
\begin{align}
\mathrm{Live}(c) &\iff \big(\exists S\in\mathcal{S}(c)\;\forall e\in S:\ \mathrm{live}(e)\big)\ \wedge\notag\\ &\qquad\ \big(\forall e\in R(c)\cup D(c):\ \mathrm{live}(e)\big),\label{eq:live}\\
\lambda(c) &= \max_{S\ \text{live}}\ \min_{e\in S\cup R(c)}\ \ell(e),\label{eq:level}
\end{align}
where $\ell$ maps proposed, accepted and provenance\_verified to $0,1,2$. The claim state is a deterministic function: \texttt{contested} if $\mathrm{Live}(c)$ and contradicting evidence is live; else \texttt{verified} ($\lambda\!=\!2$), \texttt{supported} ($\lambda\!=\!1$) or \texttt{unsupported} ($\lambda\!=\!0$); if $\neg\mathrm{Live}(c)$, \texttt{stale} whenever a registered sufficient set, premise or dependency exists, so that something once registered has been lost. A claim's level is that of the weakest member of its best live set; $\lambda$ propagates provenance tiers and has no further epistemic interpretation. A claim citing only proposed evidence is \texttt{unsupported} (it never had a live sufficient path at level $\ge1$), whereas one whose accepted evidence is later rejected is \texttt{stale}. Fig.~\ref{fig:support} shows one claim under four revisions, and Supplementary Table~S20 maps the relation vocabulary onto this structure. The earlier binary-edge semantics is the special case with only singleton sets, $R\!=\!D\!=\!\emptyset$ and no lineage.

\begin{proposition}\label{prop:conj}
Under singleton-edge semantics $\mathrm{Live}(c)=\bigvee_{e}\mathrm{live}(e)$; for no edge set is this equal to $\mathrm{live}(E_1)\wedge\mathrm{live}(E_2)$ on all assignments.
\end{proposition}
\begin{proof}
If $E_1$ is an edge, the disjunction holds when only $E_1$ is live while the conjunction fails; symmetrically for $E_2$; if neither is an edge the disjunction ignores one of them. \qedhere
\end{proof}
The result is elementary; its point is that support sets are \emph{strictly} more expressive, so losing a conjunction at registration time is a failure distinct from losing an edge.

\paragraph{What the components mean, and what they do not}
By Eqs.~\eqref{eq:live}--\eqref{eq:level}, adding the members of $R(c)$ to every set in $\mathcal{S}(c)$ leaves $\mathrm{Live}$ and $\lambda$ unchanged (only the \texttt{stale}/\texttt{unsupported} label differs for a claim with premises but no sets), so $R(c)$ adds no expressive power. It is a normalization device whose value is registration robustness: ``common to every alternative'' is declared once and its omission/flattening becomes separately measurable. $D(c)$ differs from $R(c)$ only in not entering $\lambda$. The difficult step is therefore not evaluation of Eqs.~\eqref{eq:live}--\eqref{eq:level} but \emph{construct assignment}: deciding whether a replication is required or reinforcing, whether a statistical specification is a premise, a non-epistemic dependency or part of the claim, and whether apparently independent alternatives share latent dependence. TRL takes those judgements as input. Evidential support is binary: there is no partial support, evidence-strength weighting or statistical dependence between nominally alternative sets. A manuscript assertion is a span explicitly bound to one or more claims; TRL does not segment text or decide binding granularity, so manuscript protection is conditional on binding quality. Supplement~S7 provides a counterfactual annotation codebook and preregistered agreement test; because no independent annotation round has been completed, $R(c)$, $D(c)$ and support-set membership should be read as operational categories whose inter-rater construct validity remains open.

\subsubsection{Invariants and transactions}
\label{sec:txn}
A candidate state is \emph{ledger-admissible} only if it satisfies the following invariants.
\begin{enumerate}
    \item[$I_1$] \textbf{Referential integrity:} every referenced run, evidence, claim, decision or assertion exists.
    \item[$I_2$] \textbf{Reciprocity and derivation:} every relation is recorded on both endpoints, every support-set member is indexed on its claim, and each claim and assertion state equals its derivation.
    \item[$I_3$] \textbf{Supersession and lineage closure:} superseded evidence names an existing replacement; supersession and lineage chains are acyclic.
    \item[$I_4$] \textbf{Decision provenance:} a decision has evidence or a rationale, and a decided or revised decision cites no non-live evidence (including through lineage or run invalidation).
    \item[$I_5$] \textbf{Artifact and verification integrity:} artifact digests match; evidence is not provenance-verified while a recorded tier failed or its run is unfinished.
    \item[$I_6$] \textbf{Manuscript admissibility:} manuscript-referenced claims are supported or verified; every non-outdated assertion is ledger-admissible under its registered bindings.
    \item[$I_7$] \textbf{Predecessor freshness:} the ledger head still equals the predecessor hash the transaction was built from.
    \item[$I_8$] \textbf{Transition legality:} every object moves only along its state machine (Table~\ref{tab:states}).
\end{enumerate}
$I_1$--$I_6$ and $I_8$ are deterministic checks by ordinary code; $I_7$ is optimistic concurrency control under a local commit lock. An LLM may propose a mutation but cannot override a failed check.

A transaction $\tau=(b,\Delta,a,r)$ carries the exact predecessor hash $b=h(S_t)$, operations $\Delta$, actor $a$ and rationale $r$. It is applied to a private copy, $\widehat S=\mathrm{Apply}(S_t,\Delta)$, and with $H_t$ the head at activation,
\begin{equation}
S_{t+1}=\begin{cases}\widehat S, & H_t=b\ \land\ \bigwedge_{j\in\{1,\dots,6,8\}} I_j(S_t,\widehat S)=1,\\ S_t, & \text{otherwise.}\end{cases}
\label{eq:commit}
\end{equation}
A commit writes a content-addressed snapshot $h_{t+1}=\mathrm{SHA256}(S_{t+1})$ and a hash-chained log record with actor, rationale, operations and before/after hashes. The chain is tamper-\emph{evident} against partial local edits; without an external signature an adversary with full write access could recompute it. Commit-safety (every committed state satisfies the invariants; a stale candidate cannot overwrite a newer head) follows by induction from Eq.~\eqref{eq:commit} and is stated for completeness, not as a deep result.

\begin{proposition}\label{prop:prop}
Let $G$ be the registered graph and $E_0$ the evidence rejected or superseded in one transaction. After commit every claim, assertion and decision has the state that Eqs.~\eqref{eq:live}--\eqref{eq:level} and the state machines assign on $G$ with liveness recomputed on the lineage closure of $E_0$, and nothing outside that closure's dependents changes.
\end{proposition}
\emph{Sketch.} Propagation reconciles exactly the claims linked to a closure member, the assertions bound to them and the decisions citing a member; nothing else is related to a record whose liveness changed. $I_2$ then rejects the commit unless every stored state equals its derivation. \qedhere

The result is sound and complete \emph{with respect to $G$} and silent about the true dependency structure; its content is largely definitional. What is not definitional is the choice of semantics (Proposition~\ref{prop:conj}) and how well $G$ is registered, both tested below. A replacement for superseded evidence does not inherit its support relations; they must be re-established explicitly.

\subsubsection{Dependency registration}
\label{sec:registration}
Everything above operates on the registered graph, so registration is the principal source of unprotected failure. Writing $G=\mathcal{A}(T)$ for the graph an acquisition process $\mathcal{A}$ (a stage hook, a rule, a person or an LLM) derives from a trajectory $T$, TRL specifies only the revision step applied to $G$; its guarantees (Proposition~\ref{prop:prop}) are conditional on $G$, and end-to-end protection cannot exceed the fidelity of $\mathcal{A}$. We do not implement automatic relation discovery or evaluate $\mathcal{A}$ on independently annotated real trajectories. TRL instead provides explicit registration channels---pipeline hooks called by the stage that knows the semantics, an adapter for ResearchLedger workspaces, and run capture that binds evidence to immutable runs and artifacts---and preserves registration provenance. For the confirmatory study, $G^\star$ is an \emph{adjudicated reference graph}, not assumed ground truth by fiat: blind annotator agreement is measured first, and low agreement is itself evidence that the vocabulary is not sufficiently well defined. Conditional on an adjudicated $G^\star$, we define typed precision and recall $\mathrm{P}_{\mathrm{dep}}=|G\cap G^\star|/|G|$ and $\mathrm{R}_{\mathrm{dep}}=|G\cap G^\star|/|G^\star|$ (group and alternative-group edges interchangeable), \emph{object coverage} (reference dependency-bearing objects with every incoming reference edge registered) and \emph{support-set coverage} (reference claims whose full family of sufficient sets and premises is reproduced). Two failures that $\mathrm{R}_{\mathrm{dep}}$ hides are counted separately: a \emph{flattened conjunction} (a group or premise registered as independent \texttt{supports}: no missing edge, wrong semantics) and a \emph{spurious edge} (which can disable protection by creating a fake alternative path).

\subsection{Implementation and evaluation design}
\label{sec:design}
\subsubsection{Implementation}
TRL is a model-agnostic Python layer (typed dataclasses, JSON, SHA-256, a JSONL log) around AutoResearch's 23-stage workflow; pipeline hooks record evidence, claims, decisions and manuscript bindings, and a fail-closed bridge to ResearchLedger supplies immutable runs, artifact hashing, tracing and manuscript audit (\suppl{S2}). The core specification in Table~\ref{tab:states} and the ResearchLedger v2.2 bridge use different native status vocabularies. In particular, the tracked-proxy bridge reports ResearchLedger's \texttt{supported}$\to$\texttt{hypothesis}$\to$\texttt{supported} sequence; \texttt{hypothesis} there means that no currently sufficient live support path remains and is not a Table~\ref{tab:states} TRL claim label. We retain native bridge labels rather than silently remapping them; Supplement~S2 documents the semantic correspondence. The integrated v2.2 validation build---including support-set revision, transaction recovery, tracked trajectories, paper audit and study tooling---passed 206/206 automated tests across 22 modules in non-overlapping batches; no lint or static-type result is claimed for that run because those executables were unavailable.

\subsubsection{Evidence hierarchy}
The evaluation separates four evidentiary questions that are often conflated: (i) whether the formal/state semantics are internally coherent, (ii) whether the implementation realizes those semantics, (iii) whether independent experts can map real research trajectories into the proposed dependency vocabulary, and (iv) whether using the system improves end-to-end revision behavior for real paper agents or human auditors. The current paper provides substantial evidence for (i)--(ii), bridge evidence toward (iii), and protocols rather than results for the decisive independent tests in (iii)--(iv) (Supplementary Table~S21).

\subsubsection{Confirmatory old/new paper-agent benchmark (specified, not executed)}
The scientific validation this work needs has reference judgements outside TRL. The preregistered confirmatory study targets at least 60 genuine semantic revision events from at least 15 repositories (preferred 100--150 events from 25+ repositories). For each event, the old commit is frozen, an executable paper agent is generated where feasible (Paper2Agent provides one concrete standardized substrate \cite{miao2026paper2agent}), a prespecified task is executed, and its outputs and downstream claims are persisted before the historical correction is replayed. Two annotators, blind to registrations and system predictions, independently label relations and affected claims, decisions and assertions; a third adjudicates. \emph{Construct validity is a gate, not an assumption}: agreement on edge presence, relation type and support structure is reported before system accuracy, and the preregistration treats low agreement as evidence that the relation vocabulary itself requires refinement. Primary system outcomes are stale-assertion escape and false invalidation; secondary outcomes include claim precision/recall, selective recomputation, executions avoided relative to full rebuild, recovery correctness, time-to-safe-state, dependency precision/recall and support-set coverage. The preregistered matched-registration systems are: no revision-state management; invalidate-all/full rebuild; downstream reachability; flat binary dependency propagation; AND/OR support-set provenance without transactions; and full TRL. Where feasible, an ATMS or positive-Boolean-provenance implementation is added as a strong formal comparator. Repository, not event, is the resampling unit for 95\% bootstrap intervals. The original EviGraph reconstruction remains a synthetic mechanism comparator only and is not treated as evidence about the native system. The prespecified study design is illustrated in Supplementary Fig.~S1. Protocols and the human-audit design are in \suppl{S7}; the executed 13-case tracked-proxy study below exercises the downstream ledger lifecycle but not real paper-agent generation or independent annotation.

\subsubsection{Mechanism experiments}
\textbf{Conformance.} RSM-1000 (single mutations of a 20-claim state), RSM-SEQ-2500 (25 cumulative mutations $\times$ 100 episodes), RSR-500 (stale-writer races), HT-250 (five classes of local-history corruption), RLV2-400 (seven documented ResearchLedger rule classes) and DRS-1400 (unregistered dependencies) are rerun on the revised core with their original seeds (\suppl{S3}). \textbf{SUP-300 and REG.} A generator builds trajectories with five claim templates (singleton 30\%, conjunctive 25\%, alternative groups 20\%, required premise 15\%, dependency 10\%), lineage, decisions and assertions; three evidence rejections per trajectory yield 815 events with non-empty impact (seed 61). REG corrupts the registered graph by dropped edges, flattened conjunctions or spurious edges (100 trajectories per point, seed 62). The generator authors the gold structure, so these are mechanism diagnostics whose template mix is arbitrary; only direction and mechanism are informative. \textbf{SM-CONF.} An exhaustive check of all 107 ordered state pairs, fuzzing with random legitimate operations, and fuzzing with forbidden direct writes. \textbf{Scaling.} $10^3$--$10^5$ triples at 1, 5 and 20 links per claim, with component timing.

\subsection{Use of AI-assisted manuscript tools}
An AI language model assisted with formatting and reorganizing the submitted LaTeX manuscript for this journal. It did not generate new observations, experiments or numerical results for this conversion. The authors are responsible for checking the full manuscript, citations, code, data and final disclosures.

\section*{Data availability}
Code, seeds, result files, the 13-case tracked-trajectory runner, transaction fault-injection outputs, blind annotation packets and aggregation/adjudication tooling, confirmatory preregistration, and protocols are in the accompanying v2.2 validation package (\suppl{S10}); the package is frozen by SHA-256.

\section*{Code availability}
Implementation code, integration materials and software documentation are publicly available at \url{https://github.com/raja21068/ResearchLedger}. The frozen experimental artifact identifiers are reported in Supplement~S10. The exact repository commit or release corresponding to the validation bundle and reported experiments should be identified for reproducible archival citation.

\section*{Author contributions}
To be completed and approved by all authors before submission; responsibilities cannot be inferred reliably from the source files.

\section*{Acknowledgements and funding}
Authors to confirm any acknowledgements, funding sources, grant identifiers and funder roles (or state explicitly that the study received no funding).

\section*{Competing interests}
A declaration of financial and non-financial competing interests must be reviewed and approved by all authors before submission.

\clearpage
\includepdf[pages=-,pagecommand={\thispagestyle{empty}},fitpaper=true]{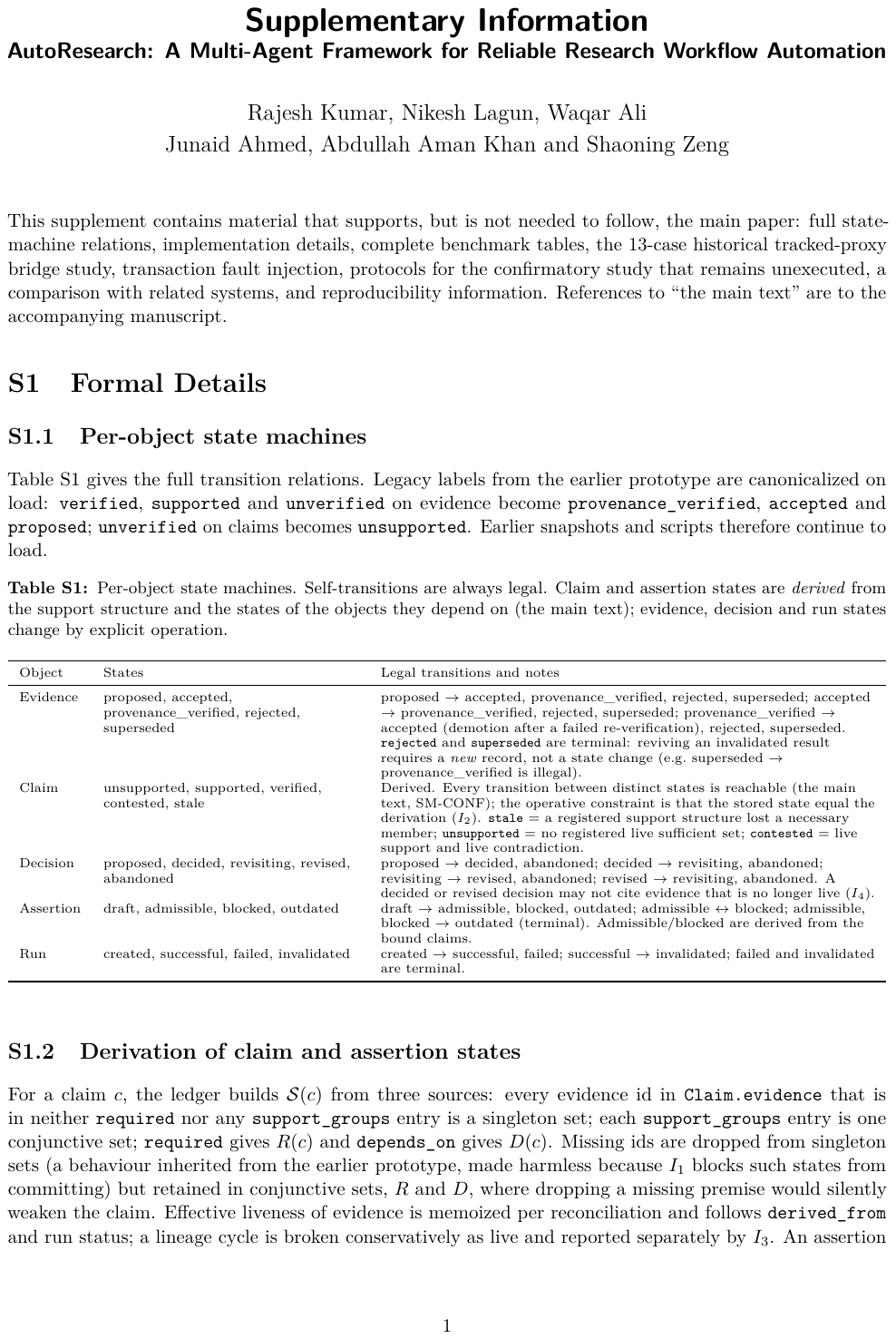}
\end{document}